\documentclass{aa}
\usepackage{graphics}
\begin{document}
 
\thesaurus{02.(02.09.1)08.(08.13.2;08.18.1)}
 
\title{Mass-loss induced
instabilities
in fast rotating stars}
\author{F. Ligni\`eres \inst{1,} \inst{2} \and C. Catala \inst{3}
\and A. Mangeney \inst{2}}
\institute{Astronomy Unit, Queen Mary \& Westfield College, Mile
End Road, London E14NS, UK \and
D\'epartement de Recherche Spatiale et Unit\'e de Recherche
associ\'ee au CNRS 264, Observatoire de Paris-Meudon, F-92195
Meudon Cedex, France \and
Laboratoire d'Astrophysique de Toulouse et Unit\'e de Recherche associ\'ee
au CNRS 285, Observatoire Midi-Pyr\'en\'ees, 14 avenue Edouard Belin, 31400
Toulouse, France}
\offprints{F. Ligni\`eres}
\mail{F.Lignieres@qmw.ac.uk}
\date{Received  ?? / Accepted ??}
\authorrunning
\titlerunning
\maketitle
\begin{abstract}
To explain the origin of Herbig Ae/Be stars activity,
it has been recently proposed
that strong mass-losses trigger 
rotational instabilities in the envelope
of fast rotating stars.
The kinetic energy transferred to turbulent motions
would then be the energy
source of the active phenomena observed
in the outer atmosphere of Herbig Ae/Be stars 
(Vigneron et al. 1990; Ligni\`eres et al. 1996).
 
In this paper, we present a one-dimensional model 
of angular momentum transport
which allows to estimate the degree of 
differential rotation 
induced by mass-loss. 
Gradients of
angular velocity are very close to $- 2 \Omega / R$ 
($\Omega$ being the surface rotation rate and $R$
the stellar radius). 
For strong mass-loss,
this process occurs in a short time
scale as compared to other processes
of angular momentum transport.
Application of existing stability criteria indicates that
rotational instabilities should develop for fast rotating star.
Thus, in fast rotating
stars with strong winds,
shear instabilities are expected
to develop and to generate
subphotospheric turbulent motions.
Albeit very simple, this model gives strong support to
the assumption made by Vigneron et al. 1990 and Ligni\`eres et al. 1996.

\end{abstract}

\section{Introduction}

Although far from complete, the overall picture describing
the angular momentum 
evolution of
solar type stars 
is well established. Despite modest mass-loss rates,
$\dot{M} = 10^{-14} {\rm M}_{\sun} {\rm yr}^{-1}$ for the sun, 
magnetised stellar winds
strongly brake the rotation of the star.
The decrease of the angular velocity then
reacts back on the magnetic field by reducing the efficiency of
the dynamo process.
This picture cannot be applied as such to early-type stars. First of all, the 
history of stellar rotation at intermediate and high masses is
generally still poorly known; besides, the presence of magnetic fields
in early-type stars has not been established, except for some categories
of chemically peculiar stars, as well as for a couple of particular cases
(Donati et al. 1997; Henrichs et al. 2000).
On the other hand, 
we know that early type stars 
can experience strong angular momentum losses
as
stellar winds with very large mass loss rate have been observed
($\dot{M} \approx 10^{-5} - 10^{-8} {\rm M}_{\sun} {\rm yr}^{-1}$). 
In this paper, we investigate how such
angular momentum losses will affect the stellar rotation, assuming
magnetic fields are not dynamically relevant.

This question has not yet received much attention although it might
have important consequences for stellar structure and evolution
as well as for the understanding of early-type stars activity.
In the context of stellar evolution, the effect of mass-loss on
rotation has to be investigated since the rotation strongly
influences the stellar structure. However, current models 
of stellar evolution with rotation do not take this effect
into account (Talon et al. 1997, Denissenkov et al. 1999). 
As explained below, we have been confronted to the same question
while investigating the origin
of the very strong activity of Herbig Ae/Be stars, a class of 
pre-main-sequence stars with masses ranging from $2$ to $5 {\rm M}_{\sun}$.

A significant fraction of these objects is known to possess extended chromospheres, 
winds, and to show high levels of spectral variability. In addition, the presence of 
magnetic fields, first suggested by the rotational modulation of certain spectral lines
(Catala et al. 1986), has been recently supported by a direct detection at the surface of 
the Herbig Ae star, HD 104237 (Donati et al. 1997).
Detailed estimates of the non-radiative heating in the outer atmosphere of Herbig Ae/Be 
stars (Catala 1989, Bouret et al. 1998) 
compared to observational constraints on the available energy 
sources strongly suggest that the rotation of the star is the only 
energy source capable of powering such activity
(see discussion in F.Ligni\`eres et al. 1996).
This led
Vigneron et al. (1990)
to propose a scenario whereby
the braking torque exerted by the stellar wind
forces turbulent motions in a differentially rotating layer
below the stellar surface. Then, by invoking
an analogy with stellar
convection zones,
these turbulent motions could generate a magnetic field
which would transfer and dissipate the turbulent kinetic
energy into the outer layers of the star.

Unlike Herbig stars, there is at present no observational evidence of
non-radiative energy input in the atmosphere of OB stars. 
Their radiatively-driven 
winds
do not require any additional 
acceleration mechanism and an eventual non-radiative heating
would be very difficult to detect because most lines are saturated.
However, they show various forms of spectral variability. 
Recent observations
seem to indicate that these phenomena
are not due to an intrinsic variability of the wind
but are instead
caused by co-rotating features on the stellar surface (Massa et al. 1995). 
As proposed in the literature (Howarth et al. 1995; Kaper et al. 1996), 
these corotating features could be due to a magnetic structuration of the wind.
Since OB stars rotate fast and possess strong winds, 
the Vigneron et al. scenario might
also explain these phenomena.
The recent direct detection of a magnetic field in an early Be star,
$\beta$~Cep, adds some credit to this hypothesis (Henrichs et al. 2000).

In this paper, we shall investigate the starting assumption of
the Vigneron et al. scenario namely that
the braking torque of an
unmagnetised wind generates strong enough velocity shear 
to trigger an instability in the subphotospheric layers.
This is a crucial step in the scenario since the onset of the instability 
allows to transfer kinetic energy from rotational motions to turbulent
motions. 

The possible connection between mass-loss and instability had been already 
suggested by 
Schatzman (1981). However, the onset of the instability is not really
considered in this study since it is assumed from the start that
the wind driven angular momentum losses induce
a turbulent flux of angular momentum. 
Here, we propose a simple model of angular momentum transport by a purely radial
mass flux in order 
to estimate the angular velocity gradients induced by mass loss.
Then, the stability of these gradients
is studied according to existing stability criteria.
Note that, as
we neglect latitudinal flows,
our model is best regarded as an equatorial model.

In the absence of magnetic fields, the braking effect
of a stellar wind is simply due to
the fact that matter
going away from the rotation
axis has to slow down to preserve its angular momentum.
Thus, for the star's surface to be significantly braked,
fluid parcels coming from deep layers inside the star have 
to reach the surface. Since radial velocities induced by
mass-loss are very small deep inside the star,
this is expected to take a relatively long time, not 
very different from the
mass-loss time scale.
By contrast, we shall see that the formation of 
unstable angular velocity gradients near the surface take place 
in a very short time, much smaller than the braking time scale.

The paper is organised as follows: first, we estimate the time
scale
characterising the braking of the stellar surface and relate it
to the mass loss time scale (Sect. 2). Then, we show that radial
outflows in stellar envelopes generate differential rotation
and we estimate the time scale of this process (Sect. 3).  
The stability of these angular velocity gradients is considered
(Sect. 4)
and the results are summarised and discussed (Sect. 5).

\section{Braking of mass-losing stars}

Generally speaking, the braking of
the star's surface
depends
on the mass-loss mechanism and on the efficiency
of angular momentum transfer inside
the star. In this section, we shall
make assumptions regarding these processes in order to estimate
the braking time scale.
However, before we consider these particular assumptions,
it is interesting to  
show that the simple fact that the star adjusts
its hydrostatic structure to its decreasing mass
already implies
that the star slows down as it loses mass.

Indeed, according to models of pre-main-sequence
evolution (Palla \& Stahler 1993),
a $2 M_{\sun}$ pre-main-sequence star
losing mass at a rate of the order of $\dot{M} =
10^{-8}  {\rm M}_{\sun} {\rm yr}^{-1}$
will have lost about one percent of its mass 
when it arrives on the main sequence.
Because mass is concentrated in the core of stars,
one percent of the total mass corresponds to 
a significant fraction of the envelope. 
For pre-main-sequence models of $2$ to $5 M_{\sun}$,
this means
that all the matter initially located above 
the radius $R_I \approx 0.63 R_*$
is expelled during the pre-main-sequence evolution.
But, during this mass-loss process, 
the star continuously adjusts its structure to its decreasing mass
and, from the point of view of stellar structure,
the loss of 1 percent of mass has a negligible effect
on the stellar radius.
Thus,
as represented on Fig.1, the sphere
containing 99 \% of the initial mass must have expanded
significantly during its pre-main-sequence evolution.
Its moment of inertia has
increased and, due to angular momentum conservation, its mean 
angular velocity has been reduced.

\begin{figure}
\resizebox{\hsize}{!}{\includegraphics{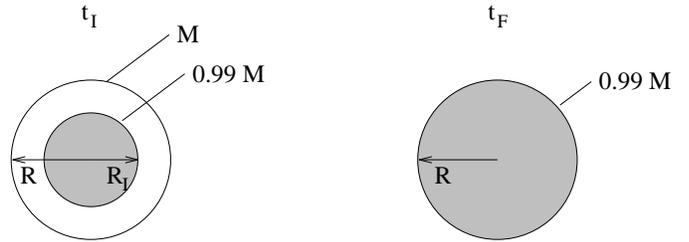}}
\caption{One percent mass loss generates an expansion of the sphere containing
$99$ percents of the stellar mass through readjustment of the stellar 
hydrostatic
structure. This expansion must be associated with a decrease of the mean
rotation rate}
\end{figure}

Thus, we conclude that the hydrodynamic adjustment of stars ensures that
mass loss is accompanied by a 
mean braking of the remaining matter.
Now, if one wants to estimate the actual braking of the stellar surface,
assumptions have to be
made on the mass-loss process and
on the efficiency of angular momentum transfer inside the star.
In order to obtain the order of magnitude of the braking time scale, 
we assume that mass-loss is isotropic and constant in time and
we
consider two extreme assumptions regarding the efficiency of angular momentum transfer.

First, we assume that the transfer of angular momentum is only due to
radial expansion. Then, 
angular momentum conservation 
states that 
the angular velocity $\Omega_I$ of a fluid parcel 
located at a radius $r=R_I$ will have decreased by $(R_I/R)^2$
when it reaches the stellar surface. This occurs when
all the matter above $r=R_I$ has been expelled so that
the decrease of the surface angular velocity can be related to the mass-loss.
We used the density structure of a $2 M_{\sun}$ pre-main-sequence model (Palla \& Stahler 1993)
for the following calculation.

Second, we assume that an unspecified transfer mechanism enforces solid
body rotation throughout the star so that angular momentum losses are 
distributed over the whole star and the braking of the stellar surface 
is less effective. In this case, 
global angular momentum conservation reads
\begin{equation}
\frac{dJ}{J} = \frac{2}{3} 
\frac{M R^2}{I} \frac{dM}{M},
\end{equation}
\noindent
where $J = I \Omega$ is the total angular momentum.
According to stellar structure models
of $2$ to $5 M_{\sun}$ pre-main-sequence stars, the 
radius of gyration,
$I / M R^2$ is close to $0.05$ so that
the 
angular velocity
decrease is approximatively given by 
\begin{equation}
\frac{\Omega_f}{\Omega_i} = \frac{I_i}{I_f} 
\left(\frac{M_f}{M_i}\right)^{40/3} 
\end{equation}
\noindent
To simply relate the braking to the mass-loss, we also assumed
that the 
decrease of the moment of inertia is proportional to the mass decrease,
or equivalently
that mass-loss induces an uniform density decrease 
throughout the star.
Albeit rough, this assumption is not critical for the conclusion drawn in 
this section. 

\begin{figure}
\resizebox{\hsize}{!}{\includegraphics{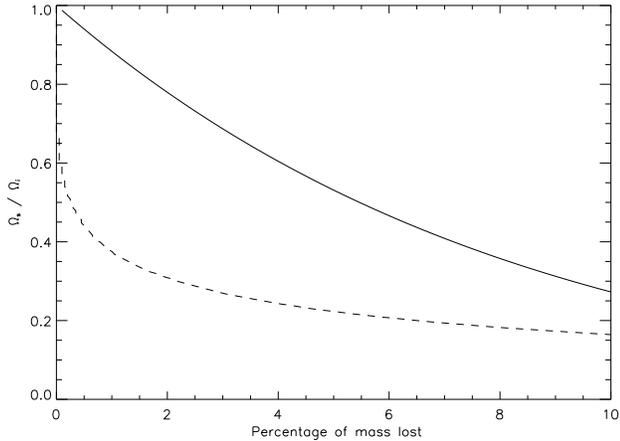}}
\caption{Braking of the stellar surface as a function of the percentage
of mass lost assuming
solid body rotation (solid line) or
purely radial expansion (dashed line)}
\end{figure}

Starting with an uniformly rotating star, 
Fig. 2 presents the braking of the stellar surface as a function
of the percentage of mass lost assuming 
solid body rotation (solid line) and 
radial
expansion (dashed line). 
This shows that, in the case of solid body rotation,
the braking time scale is about ten times smaller than the mass loss time scale,
whereas, it is hundred times smaller in the case of radial expansion 
(the braking time scale is defined as the time required to decrease the surface
angular velocity by a factor $e$). We
therefore conclude that the braking time scale is significantly smaller 
than the mass-loss time scale although both time scales
does not differ by many orders of 
magnitude.

In the next section, we study 
the formation of angular velocity gradients in a stellar envelope
assuming the transport of angular momentum is only due
to a radial mass flux. We shall see
that this process generates strong near surface gradients in a time scale 
smaller by many orders of magnitude than the mass-loss time scale.

\section{Radial advection of angular momentum across
a stellar surface}

Non-uniform radial expansion tends to generate 
differential rotation and 
this is particularly true across the stellar surface where
the steep density increase has to be accompanied by
a steep decrease of the radial velocity.
To show this let us follow the expansion of two 
spherical layers located at different depths in the star's envelope,
and rotating at the same rate.
After a given time,
the outer layer will have travelled a much larger distance
than the inner one so that its
angular velocity will have decreased 
more than 
that of the inner
layer. 
A gradient of angular velocity has then appeared between both layers.

To estimate the gradient generated by this non-uniform expansion,
we write down the angular momentum conservation assuming the
transport of angular momentum is only due to a radial flow.
Consequently, angular momentum transfers by viscous stresses, 
meridional circulation, gravity waves or turbulence are neglected all together. 
Although latitudinal inhomogeneities
in the mass-loss mechanism or Eddington-Sweet circulation
are expected to induce latitudinal flows, we note that
the present assumption would be still justified in the
equatorial plane if the meridional
circulation is symmetric with respect to the equatorial plane.

In the context of our simplified model, the angular momentum balance reads
\begin{equation}{\label{eq:momcin}}
\frac{\partial}{\partial t}(\omega)+v(r,t)
\frac{\partial}{\partial r} (\omega)=0,
\end{equation}
\noindent
where $\omega$ is the specific angular momentum and
$v(r,t)$ is the radial velocity. Up to mass loss rates
of the order of $10^{-6}  {\rm M}_{\sun} {\rm yr}^{-1}$,
radial velocities
carrying out a time-independent and isotropic mass-loss
are much smaller than the local sound speed. Consequently
the star is always very close to hydrostatic equilibrium
and continuously adjusts its structure to its decreasing mass.
This means in particular that the temporal variations
of the density near the surface are small, because the effect of 
mass-loss
is distributed over the whole star through hydrostatic equilibrium. 
Accordingly, temporal 
variations of the radial velocities
are small and we shall 
neglect them as their inclusion would
not 
modify our conclusions.
Then, the radial outflows satisfies
\begin{equation}{\label{eq:mass}}
4 \pi r^2 \varrho(r) v(r) = \dot{M},
\end{equation}
\noindent
where $\varrho(r)$ is given by stellar structure models.

With the above assumptions, the specific angular momentum evolves
like a passive scalar advected in a one dimensional stationary 
flow $v(r)$. The mathematical problem can be readily solved and we will do
so in the following for
a radial flow corresponding to
a $2 M_{\sun}$ star with
a mass loss rate of $\dot{M} =
10^{-8}  {\rm M}_{\sun} {\rm yr}^{-1}$.
But first we derive useful properties by studying the evolution 
of the angular momentum gradient in the general case.

\begin{figure}
\resizebox{\hsize}{!}{\includegraphics{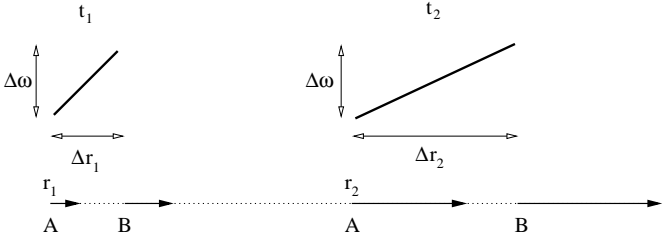}}
\caption{Gradients smoothing in an accelerated flow}
\end{figure}

Except when
$\omega$ or $v$ are uniform in space, advection
always modifies the distribution of the conserved quantity.
While decelerated flows tend to sharpen $\omega$-gradients,
accelerated flows smooth them out.
To illustrate the process of gradient smoothing in an accelerated
flow, the
angular momentum evolution of two neighbouring fluid 
elements $A$
and $B$ has been represented
in Fig.3. Following the motions, 
the angular momentum gradient between these two points decreases 
because their separation $\Delta r$ increases while 
the angular momentum difference $\Delta \omega$ is conserved.

This simple sketch can also be used to quantify the 
gradient decrease. We first note that $A$ and $B$ 
travel the distance separating $r_1 + \Delta r_1$ from $r_2$
in the same time. Thus, the time interval 
required by A to go from $r_1$ to $r_1 + \Delta r_1$ is
the same as the one used by $B$ to go from
$r_2$ to $r_2 + \Delta r_2$.
This is expressed by,
$\Delta r_2 / v(r_{2}) = \Delta r_1 / v(r_{1})$, and then,
\begin{equation} \label{eq:mi0}
v(r_{1}) \frac{\Delta \omega}{\Delta r_1} = v(r_{2})
\frac{\Delta \omega}{\Delta r_2}.
\end{equation}
\noindent
Taking the limit of vanishing separation between $A$ and $B$,
we conclude 
that the product $v(r) \partial \omega/\partial r$ is
conserved following the motions, a property which can
be readily
verified by calculating 
the Lagrangian derivative of 
$v \partial\omega/\partial r$.

This property implies that
angular momentum gradients 
can be completely smoothed out if they are advected in a 
strongly enough accelerated flow.
This is particularly relevant near stellar surfaces
where steep density gradients
induce steep radial velocity gradients.

To specify this effect we express the conservation of
$v \partial \omega / \partial r$
for the radial outflow given by Eq. (\ref{eq:mass}).
Then, the evolution of the logarithmic
gradients of the angular velocity following the flow
between radii $r_1$ and $r_2$ reads
\begin{equation}
\frac{\partial \ln \Omega}{\partial \ln r}(r_2) = -2 +
\frac{\varrho(r_2)}{\varrho(r_1)} {\left(\frac{r_2}{r_1}\right)}^{3}
\left(\frac{\partial \ln \Omega}{\partial \ln r}(r_1) +2 \right), 
\end{equation}
\noindent
where the quantity
$\frac{\partial \ln \Omega}{\partial \ln r} + 2$ measures
departures from uniform
specific angular momentum.

In this expression,
$\varrho(r) r^3$ decreases almost like $\varrho(r)$
in the vicinity of the
photosphere since the density decreases considerably
over distances small compared to the stellar radius.
Then, the above expression shows that 
any departure from constant
specific angular momentum will have decreased by a factor $e^n$
after crossing $n$ density scale heights.
Moreover, initial departures can not be too large otherwise
the corresponding differential rotation would be subjected
to powerful instabilities. Then, for any realistic values of the
initial angular velocity gradients, fluid elements reaching the surface
will have an angular velocity gradient close to $- 2 \Omega/R$.
 
We confirmed the validity of this simple picture by solving the advection
problem for the density profile corresponding to the stellar structure
model
of a pre-main-sequence 
$2 M_{\sun}$ star
(Palla \& Stahler 1993). 
Again,
the mass-loss rate
has been fixed to $\dot{M} =  
10^{-8}  {\rm M}_{\sun} {\rm yr}^{-1}$.
Starting from a solid body rotation, the evolution of the rotation profile
in the vicinity of the
stellar surface is presented in Fig.4.
The different curves correspond to increasing advection times, 
$10^{-1}$, $1$, $10$,
$10^{2}$, $10^{3}$ and $10^{4}$ years, respectively.

\begin{figure}
\resizebox{\hsize}{!}{\includegraphics{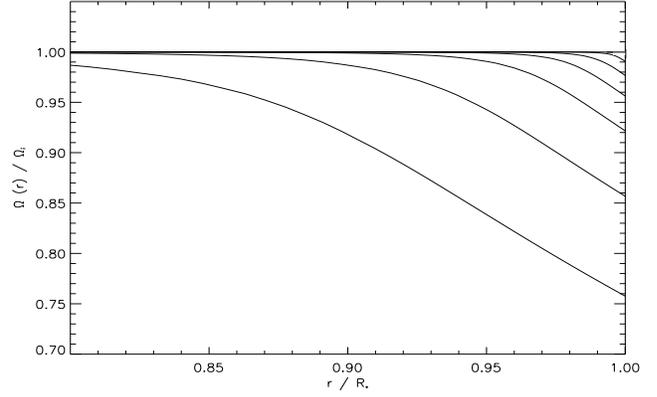}}
\caption{Rotation rate profiles
in the stellar envelope
of a 2 $M_{\sun}$ star as a result of the advection
by a radial flow with a constant mass flux 
$\dot{M} = 
10^{-8}  {\rm M}_{\sun} {\rm yr}^{-1}$.
The initial rotation rate is uniform and its evolution 
is shown after 
$10^{-1}$, $1$, $10$,
$10^{2}$, $10^{3}$ and $10^{4}$ years, respectively}
\end{figure}

We observe that the evolution is first rapid and then
much slower. This is because the evolution starts with
the rapid expansion of the outer layers
which tends to set-up a profile of 
uniform specific angular momentum
$\Omega \propto 1/ r^2$.
Once this is done,
the evolution takes place on much larger time scales
as it involves 
the much slowly expanding inner layers.
The gradual set-up of the $\Omega \propto 1/r^2$ profile is confirmed by 
the evolution of the
angular velocity logarithmic gradients. As shown on 
Fig.5, specific angular momentum gradients are rapidly
smoothed out in the outer layers while the inner ones become
affected after a long time. 

\begin{figure}
\resizebox{\hsize}{!}{\includegraphics{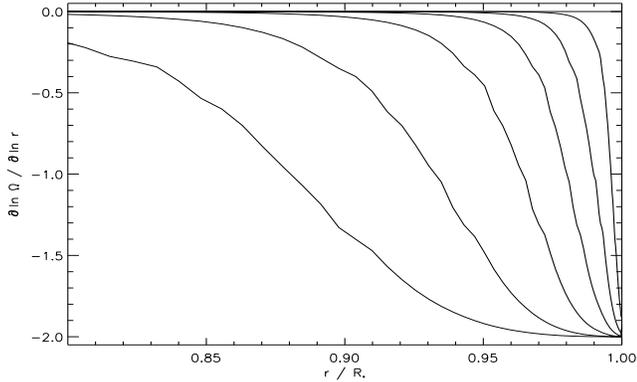}}
\caption[ ]{
Logarithmic derivative of the rotation rate. The initial derivative
vanishes uniformly 
and its evolution is shown after $10^{-1}$, $1$, $10$,
$10^{2}$, $10^{3}$ and $10^{4}$ years, respectively}
\end{figure}

We can therefore conclude that starting with
any realistic angular velocity profile
in a stellar envelope,
radial expansion will generate angular velocity gradients very close to 
$-2 \Omega / R$ near the surface.

Whether such gradients are actually present in the subphotospheric
layers of mass-losing
stars depends on how the characteristic time for the formation 
of these gradients
compares with the time
scales of other angular momentum transport processes.
The angular velocity gradients being produced by radial acceleration,
the associated time scale is
\begin{equation}
t_G = \frac{1}{\frac{\partial v}{\partial r}} = \frac{H_{\varrho}}{v(r)} =
\frac{4 \pi r^2 \varrho(r) H_{\varrho}}{M} t_{\rm M},
\end{equation}
\noindent
where $H_{\varrho}$ is the density scale height and $t_{\rm M} = M / {\dot M}$
is the mass-loss time scale.
Fig.5 shows that no more than one month is necessary to form gradients
of the order of ${\Omega} / R$ for a mass loss
rate
equal to
$\dot{M} =
10^{-8}  {\rm M}_{\sun} {\rm yr}^{-1}$. 
This time scale is smaller by many orders of
magnitude than the mass-loss
time
scale or the braking time scale. This is because the formation
of gradients requires radial displacements corresponding to
a density scale height
whereas a significant braking requires radial displacements
of the order of the stellar radius.

\section{Stability of the angular velocity gradients} 

In this section, we investigate whether the
differential rotation induced by the radial mass-loss
is sufficient to trigger a hydrodynamical instability.

The
uniform angular momentum profile being marginally stable with
respect to Rayleigh-Taylor instability, we
consider its stability with respect to shear instabilities.
When, as it is the case here,
the vorticity associated with the velocity profile does not possess 
extrema, the study of shear instabilities is much complicated by the fact
that
finite amplitude perturbations involving non-linear effects
have to be taken into account.
Existing stability criteria for such velocity profiles 
rely on laboratory experiments which
shows critical Reynolds numbers above 
which destabilization occurs.
Although the critical Reynolds number depends on the particular
flow configuration,
its value is generally of order of 
$1000$.

J.P. Zahn (1974) proposed the following instability criterion
\begin{equation} 
\frac{\partial\Omega}{\partial r} > {\left(Pr Re^ 
{\rm crit}\right)}^{1/2}\frac{N}{r}, \label{eq:stab} 
\end{equation}
\noindent
where $Re^{\rm crit}$ is the critical Reynolds number, $Pr=\nu/\kappa$
is the Prandtl number comparing the thermal diffusivity $\kappa$
to the
viscosity $\nu$ and $N$ is the Brunt-V\"{a}is\"{a}l\"{a}
frequency
which measures the strength of stable stratification in radiative interiors.
The combined effect of the stable
stratification and the thermal diffusion has been derived on
phenomenological grounds. However, the way
the criterion
depends on this effect has
been recently supported in the context of the linear stability theory
(Ligni\`eres et al. 1999).

Applying the criterion to an uniform angular momentum profile, we find that
instability occurs if the rotation period at the surface is smaller than
\begin{equation}
P \leq 18.4 \; \frac{g_{\sun}}{g} {\rm days},
\end{equation}
\noindent
where the actual temperature gradient has been taken equal to
$\nabla = 0.25$, a typical value for
radiative envelopes according to Cox (1968), and the
radiative viscosity has been assumed to dominate the molecular one
which is true in the envelope of intermediate mass stars.
Note also that this expression holds for a mono-atomic completely ionised gas 
in a chemically homogeneous
star. When radiation pressure is taken into account, the upper limit of the rotation
period has to be multiplied
by $1/\sqrt{4-3\beta}$,
where $\beta$ is the ratio between the gas pressure and the total pressure. This
factor remains very close to one for pre-main-sequence models from 
$2$ to $5 M_{\sun}$.

The above instability condition being easily met by early type stars, we 
conclude that
mass-loss tends
to impose an unstable differential rotation
below the surface of early-type stars.

\section{Discussion and conclusion}

We have studied the effect of mass-loss on the rotation of stellar envelopes 
in the simplified context of an unmagnetised spherically symmetric outflow.
In the same way as an inhomogeneous distribution
of a passive scalar
tends to be
smoothed out in a expanding flows, the radial gradients of specific
angular momentum are smoothed out by radial expansion in stellar envelopes.
This process becomes more and more effective as one approaches the surface
because expansion becomes stronger and stronger. As shown
in Figs 4 and 5, a profile
of uniform angular momentum, $\Omega \propto 1/r^2$ is rapidly
set-up in the outermost layers of the star and then
pervades towards the interior on much larger time scales.
These time scales being proportional to the mass-loss time
scale, they vary very much from stars to stars. For a typical
Herbig star ($\dot{M} =
10^{-8}  {\rm M}_{\sun} {\rm yr}^{-1}$),
an angular velocity gradient close to $-2 {\Omega}/R$ appears
in one month. By contrast, for a solar-type mass-loss rate, $\dot{M} =
10^{-14} {\rm M}_{\sun} {\rm yr}^{-1}$, it would be $10^6$ times longer
to reach the same level of differential rotation.
According to existing stability criteria, the uniform angular momentum
profile is subjected to shear instabilities provided the rotation period
is shorter than $18.4 \; g_{\sun}/g$ days.

The present one-dimensional model is admittedly
not realistic at least because
latitudinal variations occur in rotating
stars and
give rise to a meridional circulation. Nevertheless, as already mentioned,
neglecting latitudinal flows may be justified in the equatorial plane.
In the following 
we discuss to which extent the neglected processes
can prevent the
formation of turbulent differentially rotating layers
below the surface of rapidly rotating mass-losing stars.
We first consider the angular momentum transport and 
then the stability problem.

Latitudinal variations could be inherent to the mass-loss mechanism
as proposed for radiatively driven wind emitted from
fast rotating stars (Owocki \& Gayley 1997).
Such variations are
likely to generate a meridional circulation (Maeder 1999)
but,
at the latitudes where outflows occur,
we still expect radial expansion 
to play a major role in shaping the angular velocity
profile.

The Eddington-Sweet circulation driven by departure from sphericity
operates on time scale larger than the Kelvin-Helmholtz time scale of the star. 
This time scale is of the order of the order of $10^7$ years for a
$2 M_{\sun}$ Herbig stars. 
Consequently the Eddington-Sweet circulation is unlikely to prevent the formation
of the differentially rotating subphotospheric layer for strong mass-loss rate.

Turbulent motions like those generated in thermal convection zones
might prevent the formation of the differentially rotating layer.
In radiative envelopes however, it is not clear whether turbulent motions
(not generated by the mass-loss process)
would be vigorous enough.

Strong magnetic fields could in principle
prevent the formation of these gradients.
Note that
this objection is not relevant in
the context of the Vigneron et al. scenario where
the shear layer is first produced by an unmagnetised wind.

Finally, we also neglected the effect of pre-main-sequence contraction.
This is justified because
the negative radial velocities associated with
the contraction are much smaller than the positive radial
velocities induced by mass-loss in the vicinity of the stellar surface.

In what concerns the stability of the uniform angular momentum
profile, one has to remind that the stability criterion results
from an extrapolation of laboratory experiments results.
Despite recent numerical simulations which contradict
the validity of this extrapolation for a differential
rotation stable with respect to the Rayleigh-Taylor instability (Balbus et al.
1996), analysis of the experimental results
reveal that for large values of the Reynolds number
the properties of the turbulent flow do not depend on
its stability or instability with respect to the Rayleigh-Taylor criterion 
(Richard \& Zahn 1999). The solution of this debate would have to wait numerical 
simulations with higher resolution or specifically designed laboratory experiments.

The fact that the shear layer is embedded in an expanding flow may also
affect the stability because expansion is known to suppress the turbulence.
For example, this phenomenon is observed in numerical simulations
of turbulent convection near the solar surface
(Stein \& Nordlund 1998).  However, the growth time scale of the instability
being of the order of the rotation period ($\approx 1$ day for Herbig stars), 
we do not expect 
the expansion to be rapid enough to 
suppress shear turbulence.

We conclude that the present model of angular momentum advection by a radial flow
supports the assumption made by Vigneron et al. (1990) and F. Ligni\`eres et al. 
(1996) that unmagnetised winds tend to force an unstable differential rotation in the 
subphotospheric layers of Herbig Ae/Be stars. Note that this work is part of an ongoing 
effort to assess the viability of the Vigneron et al. scenario. F. Ligni\`eres et al (1996) 
and F.Ligni\`eres et al. (1998) have already studied the step further, once turbulent motions 
are generated at the top of the radiative envelope, their work pointing towards the formation 
and inwards expansion of a differentially rotating turbulent layer. In addition, 
three-dimensional numerical simulations are being performed to investigate whether 
magnetic fields can be generated in such a layer.

\begin{acknowledgements}
We are very grateful to F. Palla and S.W. Stahler for making the results
of their stellar evolution code available to us.
\end{acknowledgements}

\end{document}